\documentclass[10pt,twoside,twocolumn]{article}

\usepackage{geometry} 
\usepackage{algorithmic}
\usepackage{algorithm}
\usepackage{natbib}
\usepackage{graphicx} 
\usepackage{times}     
\usepackage[fleqn]{amsmath}
\usepackage{amssymb}
\usepackage{amsthm}
\theoremstyle{definition}
\usepackage{titlesec}
\usepackage{indentfirst}
\usepackage[font=normalfont,labelfont=it]{caption}

\usepackage{titlesec} 
\titleformat{\section}[block]{\bf}{\thesection.}{1em}{} 
\titleformat{\subsection}[block]{\it}{\thesubsection.}{1em}{} 

\newcommand{\bx}{\mathbf{x}}
\newcommand{\bX}{\mathbf{X}}
\newcommand{\bw}{\mathbf{w}}
\newcommand{\bbeta}{\boldsymbol{\beta}}
\newcommand{\bt}{\mathbf{t}}

\newcommand{\bp}{\mathbf{p}}
\newcommand{\bY}{\mathbf{Y}}
\newcommand{\by}{\mathbf{y}}
\newcommand{\bc}{\mathbf{c}}
\newcommand{\bd}{\mathbf{d}}
\newcommand{\bA}{\mathbf{A}}
\newcommand{\bb}{\mathbf{b}}
\newcommand{\bK}{\mathbf{K}}
\newcommand{\bH}{\mathbf{H}}
\newcommand{\bG}{\mathbf{G}}
\newcommand{\bI}{\mathbf{I}}

\newcommand{\bs}{\mathbf{s}}

\newcommand{\amax}{\mathop{\text{argmax}}}

\usepackage{titling} 

\title{\bf\Large On the Convergence of the Dynamic Inner PCA Algorithm} 
\author{\normalsize Sungho Shin$^\dag$, Alexander D. Smith$^\dag$, S. Joe Qin$^\ddag$, and Victor Zavala$^\dag$\thanks{To whom all correspondence should be addressed} \\[-.5ex]
  \normalsize $^\dag$University of Wisconsin-Madison Madison, WI 53705, USA.\\[-.5ex]
  \normalsize $^\ddag$University of Southern California, Los Angeles, CA 90089, USA.
}
 \date{}
\renewcommand{\maketitlehookd}{%
  \centering
  \begin{minipage}[t]{0.8\textwidth}
    \hspace{-0.3in}{\it Abstract overview}\\[1ex]
    Dynamic inner principal component analysis (DiPCA) is a powerful method for the analysis of time-dependent multivariate data. DiPCA extracts dynamic latent variables that capture the most dominant temporal trends by solving a large-scale, dense, and nonconvex nonlinear program (NLP). A scalable decomposition algorithm has been recently proposed in the literature to solve these challenging NLPs. The decomposition algorithm performs well in practice but its convergence properties are not well understood. In this work, we show that this algorithm is a specialized variant of a coordinate maximization algorithm. This observation allows us to explain why the decomposition algorithm might work (or not) in practice and can guide improvements. We compare the performance of the decomposition strategies with that of the off-the-shelf solver Ipopt.  The results show that decomposition is more scalable and, suprisingly, delivers higher quality solutions.\\

    \hspace{-0.3in}{\it Keywords}\\[1ex]
    PCA, dynamic data modeling, time series analysis, scalable
  \end{minipage}
}

\begin{document}

\maketitle

\section*{Motivation and Setting}

Principal component analysis (PCA) is a widely used method for dimensionality reduction of {\em static} multivariate data. PCA identifies latent variables that capture most information (variance) of the original data set. Dynamic inner PCA (DiPCA) is a recently proposed generalization of PCA that is used for dimensionality reduction of {\em time-dependent} data \citep{dipca}. DiPCA extracts time series for latent variables that contain most information of the original data set. DiPCA has a key advantage over augmented lagged data-based techniques, such as dynamic PCA and canonical variate analysis \citep{chiang2000fault}, in that the extracted dynamic latent variables are easy to interpret \citep{dipca}. The technique can be used in diverse application areas such as feature extraction, process monitoring, and fault detection.

\subsection*{DiPCA Formulation}
We consider time-series data $\bx_1,\bx_2,\cdots,\bx_{n+s}\in\mathbb{R}^m$ (where $m$ is the feature dimension). The data is collected in the matrix $\bX\in\mathbb{R}^{(n+s)\times m}$. We consider dynamic latent variables given by $t_i = \bw^\top \bx_i$ for $i\in\mathbb{I}_{1:n+s}$,
where $\bw\in\mathbb{R}^{m}$ is a weight vector for the latent variable subspace and $\mathbb{I}_{1:n+s}:=\{1,\cdots,n+s\}$ is a time index set. The latent variables are collected in the vector $\mathbf{t}\in\mathbb{R}^{n+s}$. We assume that the latent variables follow an autoregressive (AR) process of the form:
\begin{align}\label{eqn:lmodel}
  t_i = \beta_1 t_{i-1} + \cdots \beta_s t_{i-s} + r_i, \quad i\in\mathbb{I}_{s+1:n+s},
\end{align}
where $\bbeta\in\mathbb{R}^{s}$ is the coefficient vector for the autoregressive model, and $r_i\in\mathbb{R}$ is the residual at time $i$. We consider all vectors as column vectors and use convention $\mathbf{v}:=(v_1,\cdots,v_{n_v})$.

In DiPCA, one aims to find the weights $\bw$ and AR coefficients $\bbeta$ of the autoregressive latent variable model \eqref{eqn:lmodel} that {\em maximize the covariance} between the latent variables $t_{s+1},\cdots,t_{n+s}$ and their corresponding latent model predictions $\hat{t}_{s+1},\cdots,\hat{t}_{n+s}$, where $\hat{t}_i := \beta_{1}t_{i-1}+\cdots +\beta_{s}t_{i-s}$ for $i\in\mathbb{I}_{s+1:n+s}$. The weights and AR coefficients are found by solving an optimization problem of the form:
\begin{align}\label{eqn:dipca}
  \max_{\bw,\bbeta} \; &\sum_{i=s+1}^{n+s}t_i \hat{t}_i,\quad
  \text{s.t.}\; \Vert \bw \Vert^2_2 \leq 1,\; \Vert \bbeta \Vert^2_2 \leq 1.
\end{align}
Here, the norm constraints on $\bw$ and $\bbeta$ are used to avoid arbitrary scaling of the objective. The solution of problem  \eqref{eqn:dipca} extracts the latent variable space $\bw$ that capture the most dynamic variation in the data. With $\bw$, a subspace of latent time series that are most predictable from their past data can be obtained. One can extract all the latent time series by {\it deflating} the data matrix as $\bX\leftarrow \bX - \bt \bp^\top$ with $\bp:=X^\top \bt/\bt^\top \bt$ and by re-solving \eqref{eqn:dipca}. The last latent time series is the one that contains the least information. The whole set of latent time series can be used to reconstruct the data matrix and a subset can be used to approximate it. 

\subsection*{DiPCA Algorithm}
The DiPCA problem \eqref{eqn:dipca} is a nonconvex nonlinear program (NLP). We now analyze a decomposition algorithm (that we refer to as DiPCA algorithm I) that seeks to find solutions for this NLP. DiPCA algorithm I was proposed by \cite{dipca}. We first note that \eqref{eqn:dipca} can be expressed in the following equivalent form:
\begin{align}\label{eqn:dipca-2}
  \max_{\bw,\bbeta}\; \bw^\top \bY_{\bbeta} \bw\quad \text{s.t.}\; \Vert \bw\Vert_2^2\leq 1,\;\Vert \bbeta\Vert_2^2\leq 1,
\end{align}
where $\bY_{\bbeta}:= \sum_{i=1}^s \beta_i\bY_i$ and
\begin{align*}  
  \bY_i &:= \frac{1}{2} \left( \bX_{s+1}^\top \bX_{s+1-i} + \bX_{s+1-i}^\top \bX_{s+1} \right),\;  i\in\mathbb{I}_{1:s}\\
  \bX_{i}&:=[\bx_i\;\cdots\;\bx_{i+n-1}]^\top,\;  i\in\mathbb{I}_{1:s+1}.
\end{align*}
The algorithm aims to find a solution of the NLP by solving its first-order optimality conditions. To derive these, we note that the Lagrangian of \eqref{eqn:dipca-2} is:
\begin{align*}
  \mathcal{L}(\bw,\bbeta):=&\bw^\top\bY_{\bbeta}\bw - \lambda_w(\Vert \bw \Vert^2_2-1) - \frac{\lambda_{\bbeta} }{2}(\Vert \bbeta \Vert^2_2-1),
\end{align*}
where $\lambda_{\bbeta}$ and $\lambda_{\bw}$ are Lagrange multipliers. The first-order conditions are:
\begin{subequations}\label{eqn:kkt}
  \begin{align}
    2\bY_{\bbeta}\bw - 2\lambda_w \bw  &= 0,\; \Vert \bw \Vert^2_2=1\label{eqn:kkt-1}\\
    \bw^\top \bY_i \bw- \lambda_{\bbeta} \beta_i &= 0,\; i\in\mathbb{I}_{1:s},\; \Vert\bbeta \Vert^2_2=1.\label{eqn:kkt-2}
  \end{align}
\end{subequations}
Here, $\Vert\bw\Vert_2^2,\Vert\bbeta\Vert_2^2=1$ follow from the observation that the inequality constraints are always active. Due to nonconvexity, solving \eqref{eqn:kkt} as nonlinear equations (e.g., using Newton's method) is computationally challenging. To avoid this, the DiPCA algorithm I uses the iterative scheme:
\begin{subequations}\label{eqn:dong}
  \begin{align}
    \bw^{(\ell+1)}&= \bd^{(\ell)}/\Vert\bd^{(\ell)}\Vert_2 \label{eqn:dong-w}\\
    \bbeta^{(\ell+1)}&= \bc^{(\ell+1)}/\Vert\bc^{(\ell+1)} \Vert_2 \label{eqn:dong-b},
  \end{align}
\end{subequations}
where $\ell$ is the iteration counter and
\begin{align*}
  \bd^{(\ell)} &:= \bY_{\bbeta^{(\ell)}} \bw^{(\ell)},\;
  c_i^{(\ell)} :=(\bw^{(\ell)})^\top \bY_i\bw^{(\ell)},\; i\in\mathbb{I}_{1:s}.
\end{align*}
We observe that \eqref{eqn:kkt-1} is an eigenvalue problem and that \eqref{eqn:dong-w} attempts to approximately solve this (with fixed $\bbeta^{(\ell)}$). We will see in the next section that \eqref{eqn:dong-w} is an iteration of the so-called power method (widely used for the solution of eigenvalue problems and static PCA). We also observe that \eqref{eqn:dong-b} exactly solves \eqref{eqn:kkt-2} (for fixed $\bw^{(\ell)})$. 

The DiPCA algorithm I has shown to be rather effective at solving the first-order conditions of the NLP but no convergence guarantees have been established. Moreover, it is clear that, due to nonconvexity, the solution of the first-order conditions does not guarantee that a solution is a maximum. 

\section*{Main Results}

We now propose a coordinate maximization algorithm for solving the NLP \eqref{eqn:dipca-2} and show that a simplified variant of this approach is equivalent to the DiPCA algorithm. In coordinate maximization, one partitions the set of decision variables and solves the optimization problem over a subset of variables while fixing the rest, and repeat the procedure for each subset. This approach can be interpreted as a {\em block Gauss-Seidel} or {\em alternating maximization} scheme. Coordinate maximization is not guaranteed to converge to a local solution but is often used in applications since fixing a set of variables often reduces complexity and enables deriving closed-form solutions over the complementary variable set.  To see this, we partition the decision variables into $\bbeta$ and $\bw$. We consider solving for $\bw$ while fixing $\bbeta$:
\begin{subequations}\label{eqn:cmax}
\begin{align}\label{eqn:cmax-w}
  \bw^{(\ell+1)}=\amax_\bw\; \bw^\top \bY_{\bbeta^{(\ell)}} \bw\quad \text{s.t.}\;\Vert \bw\Vert^2_2\leq 1.
\end{align}
It is obvious that \eqref{eqn:cmax-w} is an eigenvalue problem, and such an observaion was also made in \cite[Theorem 1]{dipca}. Next, we consider solving for $\bbeta$ while fixing $\bw$:
  \begin{align}\label{eqn:cmax-b}
    \bbeta^{(\ell+1)}=\amax_{\bbeta} \; &(\bc^{(\ell+1)})^\top \bbeta
    \quad \text{s.t.}\; \Vert \bbeta \Vert^2_2 \leq 1.
  \end{align}
\end{subequations}
We observe that \eqref{eqn:cmax-b} has a closed-form solution, which is equivalent to \eqref{eqn:dong-b}. We denote \eqref{eqn:cmax} as DiPCA algorithm II. The DiPCA algorithms I and II are summarized in Algorithm \ref{alg:dipca}.

\begin{algorithm}[t!]\caption{Pseudocode for the DiPCA algorithms}\label{alg:dipca}
\begin{algorithmic}[1]
  \STATE $\ell\leftarrow 0$ and $\epsilon\leftarrow +\infty$  
  \STATE $\bY_i\leftarrow (1/2) \left( \bX_{s+1}^\top \bX_{s+1-i} + \bX_{s+1-i}^\top \bX_{s+1} \right)$
  \STATE $\by_i\leftarrow\bY_i \bw^{(0)} $ for $i\in\mathbb{I}_{1:s}$ and $\bd^{(0)} \leftarrow \sum_{i=1}^s\beta^{(0)}_i\by_i$
  \WHILE{$\epsilon>\epsilon_{\text{tol}}$}
  \STATE $\bw^{(\ell+1)}\leftarrow \bd^{(\ell)}/\Vert \bd^{(\ell)}\Vert_2$
  \STATE $\by_i\leftarrow \bY_i \bw^{(\ell+1)}$
  \STATE $c^{(\ell+1)}_i\leftarrow (\bw^{(\ell+1)})^\top \by_i$ for $i\in\mathbb{I}_{1:s}$
  \STATE $\lambda^{(\ell+1)}\leftarrow \Vert\bc^{(\ell+1)}\Vert_2$
  \STATE $^*\bbeta^{(\ell+1)}\leftarrow \bc^{(\ell+1)}/ \lambda^{(\ell+1)}$ 
  \STATE $\bd^{(\ell+1)} \leftarrow \sum_{i=1}^s\beta^{(\ell+1)}_i\by_i$
  \STATE $\epsilon\leftarrow \Vert\bd^{(\ell+1)}- \lambda^{(\ell+1)} \bw^{(\ell+1)}\Vert_\infty$ and $\ell\leftarrow \ell+1$
  \ENDWHILE
\end{algorithmic}
\footnotesize$^*$ In DiPCA algorithm II, only performed if $\lambda^{(\ell+1)}/\lambda^{(\ell)}-1<\epsilon_{\text{tol}}$.
\end{algorithm}

Solving \eqref{eqn:cmax-w} is equivalent to finding the dominant eigenvector of $\bY_{\bbeta^{(\ell)}}$. Solving \eqref{eqn:cmax-w} via an eigenvalue decomposition is computationally inefficient if the feature space $m$ is large. A more scalable approach to find the dominant eigenvector is known as the {\em power method}. The power iteration for finding the dominant eigenvalue of matrix $\bA$ is given by:
\begin{align}\label{eqn:powiter}
  \bb^{(k+1)} = \bA\bb^{(k)} / \Vert \bA \bb^{(k)} \Vert_2,\quad k=1,2,\cdots,
\end{align}
where $k$ is the iteration counter. With \eqref{eqn:powiter}, $\bb^{(k)}$ geometrically converges to the dominant eigenvector of $\bA$ if $|\lambda_1(\bA)|>|\lambda_2(\bA)|$ and $\bb^{(0)}$ has a nonzero component of the dominant eigenvector, where $\lambda_1(\cdot)$ and $\lambda_2(\cdot)$ denote the largest and the second largest eigenvalues.

The above derivations reveal connections between DiPCA algorithm I \eqref{eqn:dong} and II \eqref{eqn:cmax}. One can see that \eqref{eqn:dong-b} and \eqref{eqn:cmax-b} are identical, but \eqref{eqn:dong-w} and \eqref{eqn:cmax-w} are not. Rather, \eqref{eqn:dong-w} performs {\em one iteration of the power method} \eqref{eqn:powiter} with matrix $\bA=\bY_{\bbeta^{(\ell+1)}}$. We can also interpret \eqref{eqn:dong-w} as solving \eqref{eqn:cmax-w} with a {\em linearized} objective function. This approach is advantageous in computational efficiency since it avoids performing multiple power iterations to solve the eigenvalue problem. On the other hand, performing one iteration of \eqref{eqn:powiter} may not guarantee the improvement of the objective value and thus DiPCA algorithm I might face convergence issues, especially when the dominant eigenvalue is negative. Note that DiPCA algorithm I and II do not require performing matrix factorizations. 

The above derivations also reveal metrics for monitoring convergence. By evaluating the residual $\bs^{(\ell)}$ of \eqref{eqn:kkt-1} at iteration $\ell$ with $\lambda^{(\ell)}:=(\bw^{(\ell)})^\top \bY_{\bbeta^{(\ell)}} \bw^{(\ell)}$, we obtain $\bs^{(\ell)} := \bd^{(\ell)}- \lambda^{(\ell)} \bw^{(\ell)}$.
We also note that \eqref{eqn:kkt-2} is automatically satisfied. Consequently, one can stop the algorithm if $\Vert \bs^{(\ell)} \Vert_\infty < \epsilon_{\text{tol}}$, for user-defined tolerance $\epsilon_{\text{tol}}$. 

At a fixed point $(\bw,\bbeta)$ of iteration \eqref{eqn:dong}, one can show that the first-order conditions \eqref{eqn:kkt} hold with $\lambda_{\bw}=\lambda_{\bbeta}=\bw^\top\bY_{\bbeta}\bw$. The second-order  conditions hold (the fixed point is a maximum point) if the reduced Hessian $\mathbf{Z^\top HZ}$ is negative definite; this is equivalent to the condition that the inertia of $\bK$ is $(n_{+},n_{-},n_0)=(2,m+s,0)$, where 
\begin{align*}
  \bK&:=\begin{bmatrix} \bH& \bG^\top\\\bG \end{bmatrix},\;
  \bG:=\begin{bmatrix} \bw^\top\\&\bbeta^\top \end{bmatrix},\;\lambda:=\bw^\top\bY_{\bbeta}\bw 
  \\
  \bH&:=\begin{bmatrix}
  \bY_{\bbeta} - \lambda \bI &\bY_1\bw &\cdots &\bY_s \bw\\
  \bw^\top\bY_1& -\frac{1}{2}\lambda\\
  \vdots&&\ddots&\\
  \bw^\top \bY_s &&&-\frac{1}{2}\lambda\\
  \end{bmatrix},
\end{align*}
and $\mathbf{Z}$ is a null-space basis matrix of $\bG$.
\section*{Numerical Experiments}

We compare the performance of the DiPCA algorithm I and II with that of the off-the-shelf nonlinear programming solver Ipopt \citep{Wachter2006implementation}. Our benchmarks consist of $20$ time series obtained from data for a chemical sensor with $m=5106$, $n=71$, $s=4$ \citep{cao2018machine}. We add artificial noise to the data with iid Gaussian random variables $N(0,\sigma^2)$ and use $\epsilon_{\text{tol}}=10^{-6}$. The code is implemented in Julia and run on a Intel(R) Xeon(R) CPU E5-2698 v3 @ 2.30GHz. For Ipopt, we solve the NLP \eqref{eqn:dipca} in the space of $\mathbf{t},\bbeta,\bw$.

\begin{figure}[t!]
  \centering
  \includegraphics[width=.45\textwidth]{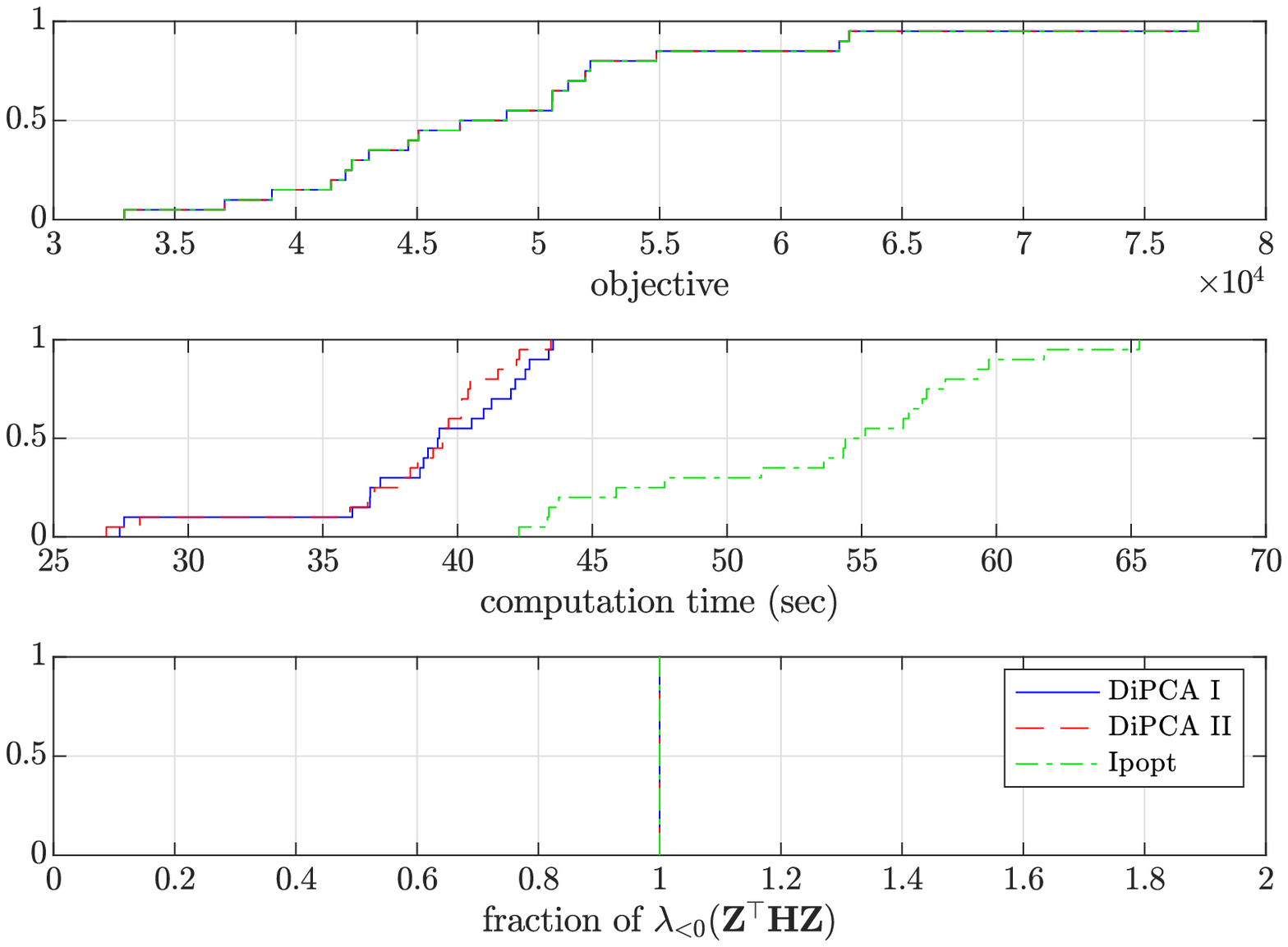}\\
  \caption{\it Benchmark results with $\sigma=1$.}\label{fig:benchmark-1}
  \includegraphics[width=.45\textwidth]{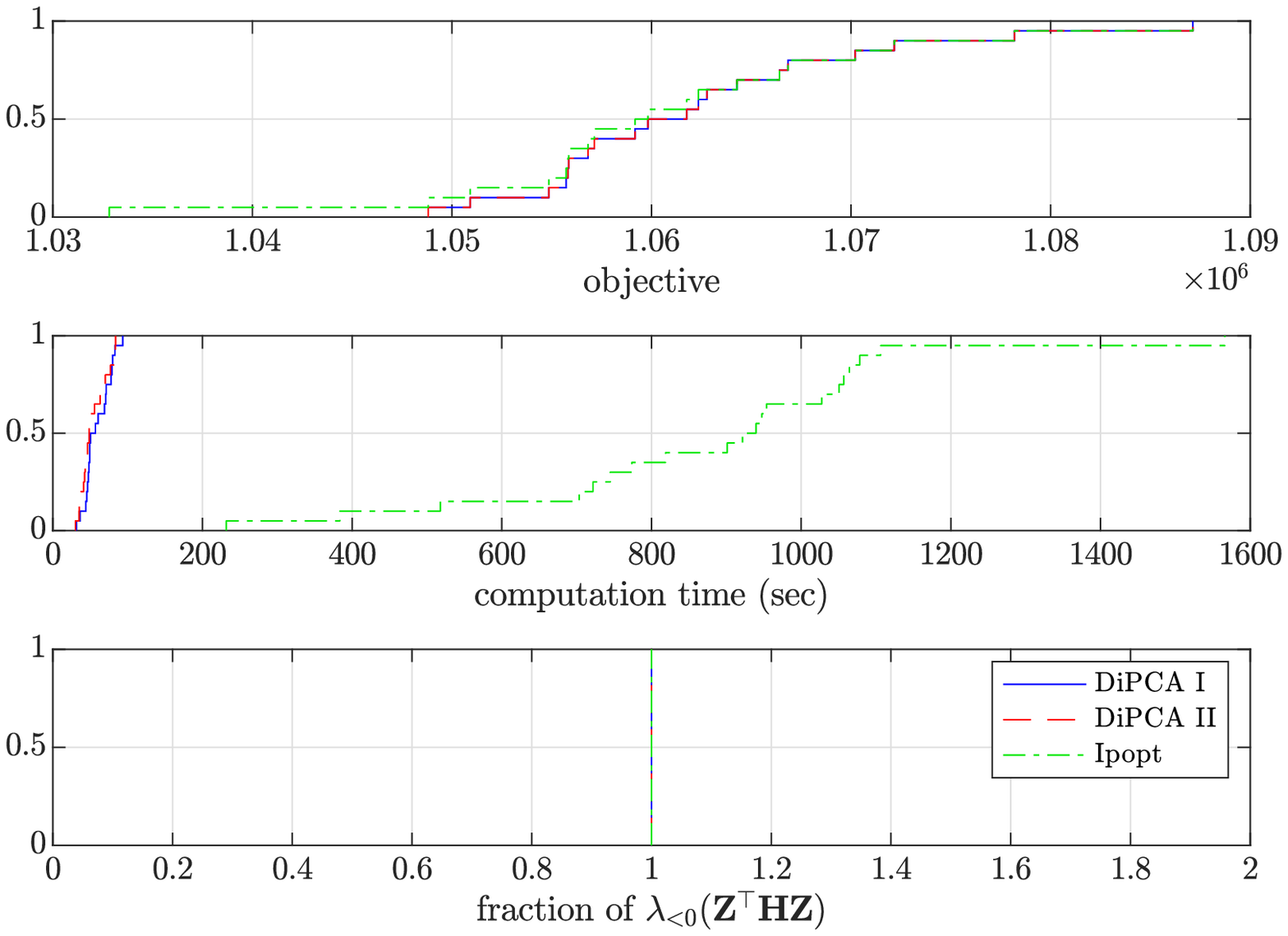}
  \caption{\it Benchmark results with $\sigma=10$ (right).}\label{fig:benchmark-2}
\end{figure}

The results are shown in the form of cumulative plots for objective value, computational time, and the fraction of negative eigenvalues among the eigenvalues of the reduced Hessian (Figure \ref{fig:benchmark-1}-\ref{fig:benchmark-2}). One can see that the performance of DiPCA algorithm I and II is similar. When the noise is small ($\sigma=1$), we can see that the DiPCA algorithms significantly outperform Ipopt in terms of computation time, but their performance is very similar in terms of objective values. When the noise is large ($\sigma=10$), the computation time drastically increases for both approaches (DiPCA and Ipopt), but DiPCA is significantly faster. The superior efficiency of DiPCA is due to the fact that Ipopt needs to perform matrix factorizations (while DiPCA algorithms does not). We have found that high noise adversely affects the conditioning of the problem (matrix $\mathbf{Y}_{\bbeta}$) and this slows down the convergence (the power iteration becomes less efficient). In the high noise case, we also found that the DiPCA algorithms find better solutions (compared to Ipopt) in terms of objective values. This seems to indicate that coordinate maximization handles nonconvexity better. 

\bibliographystyle{apa}
\bibliography{fopam}

\begin{thebibliography}{}

\bibitem[\protect\astroncite{Cao et~al.}{2018}]{cao2018machine}
Cao, Y., Yu, H., Abbott, N.~L., and Zavala, V.~M. (2018).
\newblock Machine learning algorithms for liquid crystal-based sensors.
\newblock {\em ACS sensors}, 3(11):2237--2245.

\bibitem[\protect\astroncite{Chiang et~al.}{2000}]{chiang2000fault}
Chiang, L.~H., Russell, E.~L., and Braatz, R.~D. (2000).
\newblock {\em Fault detection and diagnosis in industrial systems}.
\newblock Springer Science \& Business Media.

\bibitem[\protect\astroncite{Dong and Qin}{2018}]{dipca}
Dong, Y. and Qin, S.~J. (2018).
\newblock A novel dynamic pca algorithm for dynamic data modeling and process
  monitoring.
\newblock {\em Journal of Process Control}, 67:1--11.

\bibitem[\protect\astroncite{W{\"a}chter and
  Biegler}{2006}]{Wachter2006implementation}
W{\"a}chter, A. and Biegler, L.~T. (2006).
\newblock On the implementation of an interior-point filter line-search
  algorithm for large-scale nonlinear programming.
\newblock {\em Mathematical programming}, 106(1):25--57.

\end{thebibliography}

\end{document}